\documentclass[10pt,letterpaper]{article}
\usepackage{opex3} 

\usepackage{graphicx}

\begin{document}

\title{Detection of multimode spatial correlation in PDC and application to the absolute calibration of a CCD camera}

\author{Giorgio Brida, Ivo Pietro Degiovanni, Marco Genovese,  Maria Luisa Rastello , Ivano Ruo-Berchera\textsuperscript{*}}
\address{Istituto Nazionale di Ricerca Metrologica, Strada delle Cacce 91, 10135 Torino, Italy}
\email{$^*$i.ruoberchera@inrim.it}

\begin{abstract}
We propose and demonstrate experimentally a new method based on the
spatial entanglement for the absolute calibration of analog
detector. The idea consists on measuring the sub-shot-noise
intensity correlation between two branches of parametric down
conversion, containing many pairwise correlated spatial modes. We
calibrate a scientific CCD camera and a preliminary evaluation of
the statistical uncertainty indicates the metrological interest of
the method.
\end{abstract}



\section{Introduction}

The possibility of realizing sub shot noise regime by exploiting
multimode spatial correlation at the quantum level \cite{Kol2007,
BrambPRA2004}, sometimes called spatial entanglement, has been
experimentally demonstrated by using traveling wave parametric
amplifier and CCD array detectors both at the single photon level
\cite{BlanchetPRL2008} and at hight photon flux \cite{JedrkPRL2004,
BridaPRL2009, BridaNPHOT2010,mat,mc}, and by exploiting four wave
mixing in rubidium vapors \cite{BoyerPRL2008}. Very recently,
following the proposal of \cite{BrambPRA2008} our group showed that
it can find a natural application to the quantum imaging of weak
absorbing  object beyond the shot-noise-level
\cite{BridaNPHOT2010}(standard quantum limit). All these findings
indicate that multimode spatial correlations and their detection is
a mature field that can lead to other interesting applications. In
particular, in this framework we developed the capability of an
almost perfect spatial selection of modes that are pairwise
correlated at the quantum level in parametric down conversion, and
the optimization of the noise reduction below the shot-noise-level.

On the other side, quantum correlations of twin beams generated by
Parametric Down Conversion (PDC), is a well recognized tool for
calibrating single photon detectors, competing with the traditional
ones of the optical metrology
\cite{ge,PolMig2007,bp1,Burnham,klysh,malygin,Ginzburg,alan,n1,n2,n3,n4}.
Extending this technique to higher photon fluxes, for calibrating
analog detectors, may have great importance in metrology
\cite{BridaJOSAB2006, BridaOE2008, BridaJMO2009, RuoASL2009,
LindAO2006,Masha}. So far, the main problem to achieve this goal has
been the difficulty of an accurate spatial selection of the
correlated modes among the spatially broadband emission of PDC.

In the work presented here, the know-out in the detection of spatial correlation in PDC is fruitfully applied to the absolute calibration of CCD cameras. The method is based on the measurement of the degree of correlation between symmetric areas belonging to the twin beams, which in principle depends only by the transmission and detection efficiency. The achieved statistical uncertainty indicates the effectiveness of the methods for metrology applications.

\section{Multimode spatial correlation in PDC}\label{Multimode correlation}

\begin{figure}[tbp]
\par
\begin{center}\label{Correlations}
\includegraphics[angle=0, width=12cm, height=6 cm]{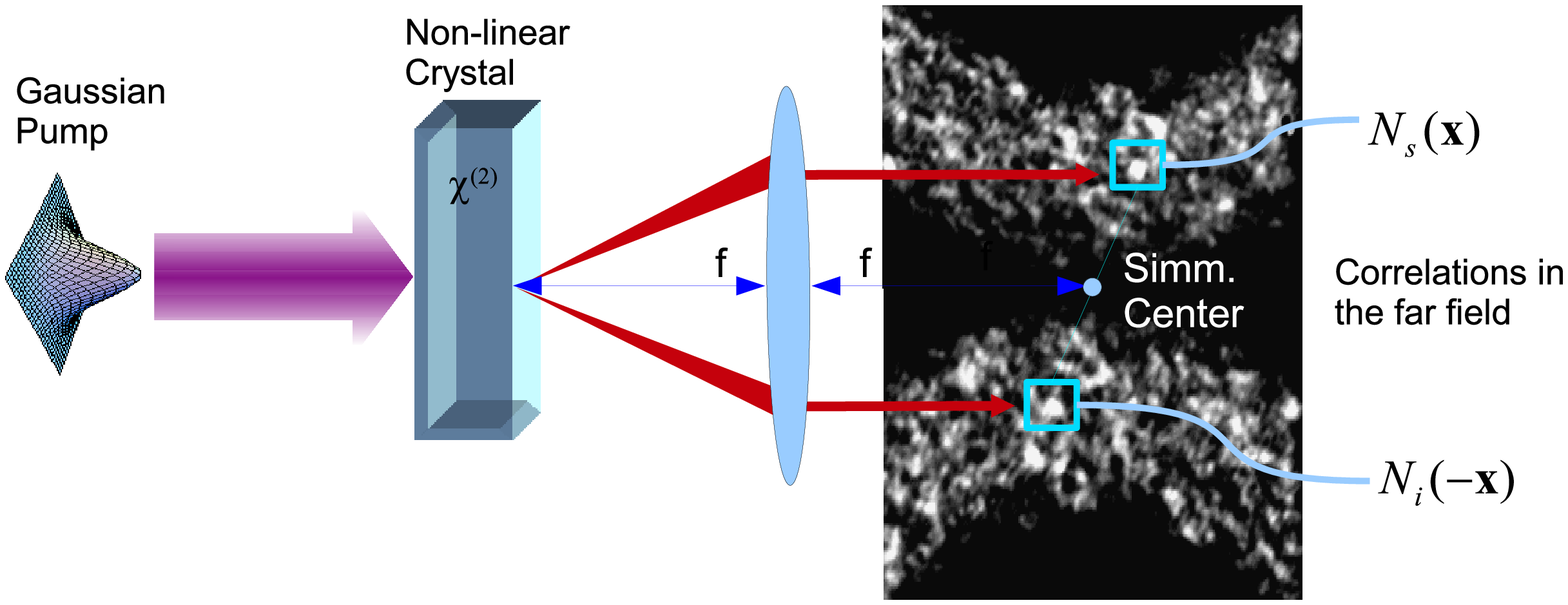}
\caption{\textsf{Scheme for spatial correlations detection in PDC.
The PDC emission from the non-linear crystal is detected in the far
field, reached by a optical $f-f$ configuration. The pump transverse
size determines an uncertainty in the photon propagation direction.
The speckled image showed, has been obtained experimentally in very
high gain regime. For this reason, the spatial fluctuations are very
strong, and the coherence area is roughly represented by the typical
size of the speckles. }}
\end{center}
\end{figure}

The state produced by spontaneous PDC, in the approximation of plane wave pump field of frequency $\omega_{p}$ and propagating
in the $z$ direction, presents perfect transverse momentum phase-matching. Thus, it can be expressed as a tensor product of
bipartite states univocally identified by frequency $\omega_{p}/2\pm\omega$ and transverse momentum $\pm\mathbf{q}$, i.e.
$|\Psi\rangle=\bigotimes_{\mathbf{q},\omega}|\psi(\mathbf{q},\omega)\rangle$. Since we are mainly interested to the frequencies
near to the degeneracy ($\omega\sim0$), the state of the single bipartite transverse mode reduces to
\begin{equation}\label{two-mode state}
|\psi(\mathbf{q})\rangle=\sum_{n}C_{\mathbf{q}}(n)|n\rangle_{i,\mathbf{q}}|n\rangle_{s,-\mathbf{q}},
\end{equation}
where the coefficients $C_{\mathbf{q}}(n) \propto \sqrt{\langle n_{\mathbf{q}}\rangle^{n}/\langle
n_{\mathbf{q}}+1\rangle^{n+1}}$ are related to the mean number of photon in the mode $\mathbf{q}$ assumed to be the same for all
the modes, i.e. $\langle n_{\mathbf{q}}\rangle = \mu$, and the subscripts $s$ and $i$ indicated signal and idler fields.

The two-mode state in Eq. (\ref{two-mode state}) is entangled in the number of photons for each pair of modes $\pm\mathbf{q}$,
whereas the statistics of the single mode is thermal with mean value $\langle n_{i,\mathbf{q}}\rangle = \langle
n_{s,\mathbf{-q}}\rangle = \mu$ and variance $\langle \delta^{2}n_{i,\mathbf{q}}\rangle=\langle
\delta^{2}n_{s,\mathbf{-q}}\rangle=\mu (\mu+1)$.  Now, we focus on the far field region, obtained as the focal plane of a thin
lens of focal length $f$ in a $f-f$ configuration (Fig. \ref{Correlations}). Here, any transverse mode $\mathbf{q}$ is
associated with a single position $\mathbf{x}$ in the detection (focal) plane according to the geometric transformation $(2c
f/\omega_{p})\mathbf{q}\rightarrow \mathbf{x}$, with $c$ the speed of light. Therefore, a perfect correlation appears in the
photon number $n_{i,\mathbf{x}}$ and $n_{s,-\mathbf{x}}$ registered by two detectors placed in two symmetrical positions
$\mathbf{x}$ and $-\mathbf{x}$, where the center of symmetry (CS) is basically the pump-detection plane interception (neglecting
the walk off).

In real experiments the pump laser is not a plane wave, rather it
can be reasonably represented by a gaussian distribution with
spatial waist $w_p$. This induces an uncertainty in the relative
propagation directions of the twin photons of the order of the
angular bandwidth of the pump. This uncertainty is the coherence
area of the process, roughly corresponding to the transverse size of
the mode in the far field (see Fig. \ref{Correlations}). The number
of photons collected in symmetrical portions of the far-field zone
are perfectly quantum correlated only when the detection areas
$\mathcal{A}_{det}$ are broader than a coherence area, whose size
$\mathcal{A}_{coh}$ is of the order of $[(2\pi c
f)/(\omega_{p}w_{p})]^{2}$ \cite{BrambPRA2004,BrambPRA2008} (only
for very large parametric gain it deviates from this behavior
\cite{BrambPRA2008,JedrkPRL2004,BGMPR-IGQI2009,BGMPR-JMO2009}).
Therefore, let us consider to collect photons over two perfectly
symmetrical and correlated areas $\mathcal{A}_{det,s}$ and
$\mathcal{A}_{det,i}$ with detection efficiency $\eta_{s}$ and
$\eta_{i}$ belonging to the signal and idler beams respectively, and
containing many transverse spatial modes
$\mathcal{M}_{spatial}=\mathcal{A}_{det,j}/\mathcal{A}_{coh}$
($j=s,i$). We also consider a situation in which the detection time
$\mathcal{T}_{det}$ is much larger than the coherence time
$\mathcal{T}_{coh}$of the process, thus the number of temporal mode
is large, $\mathcal{M}_{t}=\mathcal{T}_{det}/\mathcal{T}_{coh}\gg1$.
Since the modes in the single region are independent, the statistics
is multithermal with mean value $\langle
N_{j}\rangle=\mathcal{M}_{tot} \eta_{j} \mu$ (with
$\mathcal{M}_{tot}=\mathcal{M}_{t}\mathcal{M}_{spatial} $) and
variance
\begin{equation}\label{EN}
\left\langle\delta ^{2}N_{j}\right\rangle\equiv\left\langle N_{j}\right\rangle\left(1+\mathcal{E}\right)=\left\langle
N_{j}\right\rangle\left(1+\frac{\left\langle N_{j}\right\rangle}{\mathcal{M}_{tot}}\right)= \mathcal{M}_{tot}\eta_{j}
  \mu  \left(1+\eta_{j}\mu\right),
\end{equation}
with $j=s,i$ and $\mathcal{E}$ the excess noise, usually defined as the fluctuations that exceed the shot noise level (SNL). The
SNL, or standard quantum limit  represents the level of noise associated to the detection of the coherent light, i.e.
$\langle\delta ^{2}N_{j}\rangle_{SNL}=\langle N_{j}\rangle$. In this theoretical description, the excess noise is only related
to the intrinsic thermal statistic of the single beam of PDC  $\mathcal{E}=\langle N_{j}\rangle/\mathcal{M}_{tot}$. However, we
will discuss in the following that experimental imperfections give the major contribution to the excess noise in our setup.

The covariance between the signal and idler numbers of photon is
\begin{equation}\label{correlation}
\left\langle\delta N_{i}\delta N_{s}\right\rangle=\mathcal{M}_{tot}\eta_{s}\eta_{i} \mu(1+ \mu).
\end{equation}
The amount of correlation between the signal and idler fields is usually expressed in terms of the noise reduction factor $\sigma$
defined as the fluctuation of the difference $N_{-}\equiv N_{s}-N_{i}$ between the photons number normalized to the
corresponding level of shot noise:
\begin{equation}\label{sigma}
\sigma\equiv\frac{\left\langle\delta ^{2}N_{-}\right\rangle} {\left\langle N_{i}+N_{s}\right\rangle}=
1-\eta_{+}+\frac{\eta_{-}^{2}}{2\eta_{+}}\left(\frac{1}{2}+ \mu \right)=
1-\eta_{+}+\frac{\eta_{-}^{2}}{4\eta_{+}^{2}}\left(\eta_{+}+\frac{\langle N_{s}+N_{i}\rangle}{\mathcal{M}_{tot}}\right),
\end{equation}
where $\eta_{+}=(\eta_{s}+\eta_{i})/2$ and $\eta_{-}=\eta_{s}-\eta_{i}$. It as been evaluated by introducing the Eq.s
(\ref{correlation}) and (\ref{EN}) in the expression of the fluctuation $\langle\delta ^{2}N_{-}\rangle\equiv\langle\delta
^{2}N_{s}\rangle+\langle\delta ^{2}N_{i}\rangle-2\langle\delta N_{i}\delta N_{s}\rangle$

In the case of perfect  balanced  losses $\eta_{s}=\eta_{i}=\eta$ one get that $\sigma=1-\eta$ only depending on the quantum
efficiency. Therefore, in an ideal case  in which $\eta\rightarrow 1$, $\sigma$ approaches zero. On the other side, for
classical states of light the degree of correlation is bounded by $\sigma\geq1$,  where the lowest limit is reached for coherent
beams, $\sigma=1$.

According to Eq. (\ref{sigma}), the absolute estimation of the quantum efficiency by measuring the noise reduction factor, can
be achieved just when the excess noise disappears, that is realized in the case of balanced losses. The balancing of the two
channels can be performed physically by adding proper absorbing filters in the optical paths. Anyway, a more convenient approach
is to compensate a posteriori, for instance multiplying the values of $N_{i}$ for a factor $\alpha=\left\langle
N_{s}\right\rangle/\left\langle N_{i}\right\rangle=\eta_{s}/\eta_{i}$. It corresponds to evaluate the redefined noise reduction
factor $\sigma_{\alpha}$ (instead of the one in Eq. (\ref{sigma})):
\begin{equation}\label{sigma_alfa}
\sigma_{\alpha}\equiv\frac{\left\langle\delta ^{2}(N_{s}-\alpha N_{i})\right\rangle} {2\left\langle
N_{s}\right\rangle}=\frac{1}{2}(1+\alpha)-\eta_{s},
\end{equation}
This relation shows that the quantum efficiency $\eta_{s}$ can be evaluated measuring $\sigma_{\alpha}$ and the ratio
$\alpha=\left\langle N_{s}\right\rangle/\left\langle N_{i}\right\rangle$, without the need of a physical balancing the two
optical paths.

As a final remark, we observe that the result in Eq.
(\ref{sigma_alfa}) relays on the assumption that each spatial modes
collected by the region $\mathcal{A}_{det,i}$ finds its correlated
in the region $\mathcal{A}_{det,s}$, and viceversa. Otherwise, the
presence of uncorrelated modes in the two region would not provide a
complete cancelation of the excess noise by the subtraction, leading
to underestimate the quantum efficiency \cite{Chekhova2010}.
Therefore, experimental control on the detection of the spatial
modes by means of precise positioning and sizing of the regions and
accurate determination of the CS is fundamental for the accuracy of
the estimation of the quantum efficiency.

\section{The experimental procedure}
In our setup, a type II BBO non-linear crystal ($l=7$ mm) is pumped by the third harmonic (355 nm) of a Q-switched Nd:Yag laser.
The pulses have a duration of $\mathcal{T}_p=5$ ns with a repetition rate of 10 Hz and a maximum energy, at the selected
wavelength, of about 200 mJ. The pump beam crosses a spatial filter (a lens with a focal length of 50 cm and a diamond pin-hole,
250 $\mu$m of diameter), in order to eliminate the non-gaussian components and then it is collimated by a system of lenses to a
diameter of $w_{p}=1.25$ mm. After the crystal, the pump is stopped by a couple of UV mirrors, transparent to the visible
($\simeq 98\%$ transmission at 710 nm), and by a low frequency-pass filter ($\simeq 95\%$ transmission at 710 nm). The down
converted beams (signal and idler) are separated in polarization by two polarizers ($97\%$ transmission) and finally the far
field is imaged by a CCD camera. We used a 1340X400 CCD array, Princeton Pixis:400BR (pixel size of 20 $\mu$m), with high
quantum efficiency (around 80\%) and low noise in the read out process ($4$ electrons/pixel). The CCD exposure time is set by a
mechanical shutter to 90 ms, thus each image acquired corresponds to the PDC emission generated by a single shot of the laser.
The far field is observed at the focal plane of the lens with 10 cm focus in a $f-f$ optical configuration.

For reasons related to the visibility of the correlation, and in order to reduce the contribution of the read noise of the CCD,
it is convenient to perform an hardware binning of the physical pixels. It consists in grouping the physical pixels in squared
blocks, each of them being processed by the CCD electronics as single "superpixel". Depending on the measurement, the size of
the superpixel can be set accordingly. Typically we choose it of the same order, or larger than $\mathcal{A}_{coh}$.

The expected number of temporal modes $\mathcal{M}_{t}=\mathcal{T}_{p} / \mathcal{T}_{coh}$ detected in one image is $5\cdot
10^{3}$, considering the coherence time $\mathcal{T}_{coh}$ of PDC around one picosecond. The number of spatial modes
$\mathcal{M}_{spatial}=\mathcal{A}_{det,j}/\mathcal{A}_{coh}$ ($j=s,i$) depends only on the size of the detection areas, since
$\mathcal{A}_{coh}\sim 120\times120(\mu m)^{2}$ is fixed by the pump transverse size. The level of excess noise due to the
thermal statistics in the single beam is also fixed, since we keep fixed (aside unwanted fluctuation pulse-to pulse) the power
of the laser. The total number of modes $\mathcal{M}_{tot}$ turn out to be compatible with the level of excess noise
$\mathcal{E}\equiv\left\langle N_{s}\right\rangle/\mathcal{M}_{tot}\sim 0.1-0.2$. It can be measured by performing spatial
statistics as described in Sec. (\ref{evaluation of the CS}).

For an accurate estimation of quantum efficiency the following steps should be performed:
\begin{itemize}
\item [a)]\textit{Determination of the center of symmetry (CS)} \\ positioning of the correlated areas and determination of the center of symmetry of the spatial correlations within sub-coherence-area uncertainty, according to the experimental procedure presented in Subsection (\ref{evaluation of the CS})
  \item [b)]\textit{Determination of the minimum size of $\mathcal{A}_{det,j}$}\\
  The size of the detection areas must satisfy the condition   $\mathcal{M}_{spatial}=\mathcal{A}_{det,j}/\mathcal{A}_{coh}\gg1$, for the purposes of the unbiased estimation of $\sigma$.
  This can be achieved following the procedure sketched in Subsection (\ref{A_det aetimation}).
  \item [c)]\textit{Analysis of experimental contributions to the excess noise}\\
  Evaluation of noise coming from experimental imperfections, such as instability of the laser pulse-to-pulse
   energy and the background due to straylight and electronic noise of the CCD. Eq.s (\ref{alfaE}) and
   (\ref{sigmaalfaE}) are modified in order to account for these noise contributions. Detailed discussion on this item can be found in
   Subsection (\ref{Contributions to the excess noise}).
   \item [d)]\textit{Estimation of $\eta_j$ and of its statistical uncertainty according to Eq. (\ref{sigma_alfa})}\\
   $\alpha$ and $\sigma_{\alpha}$ are estimated experimentally over a set of $\mathcal{N}$ images according to formula
\begin{equation}\label{alfaE}
\alpha=\frac{E[N_s (k)]}{E[N_i (k)]},
\end{equation}
and formula
\begin{equation}\label{sigmaalfaE}
\sigma_{\alpha}=\frac{E[(N_{s} (k)- \alpha N_{i} (k) )^2]-E[N_{s} (k)- \alpha N_{i} (k)]^2 }{E[N_s (k)]+\alpha E[N_i (k)]},
\end{equation}

 respectively, where $E[N_j(k)]=\mathcal{N}^{-1} \sum_{k=1}^{\mathcal{N}}N_j(k)$, and $N_j(k)$ is the number of photons observed in
 the detection
area $\mathcal{A}_{det,j}$ in the $k$-th image. [See Subsection (\ref{Efficiency estimation and uncertainty})].
   \item [e)]\textit{Evaluation of the optical losses}\\
   The actual value for the estimated quantum efficiency $\eta_j$ subsumes also losses due to the crystal, the lens and the mirrors.
   Thus, the value of the quantum efficiency of the CCD is obtained as the ratio between the estimated value for the quantum efficiency
   $\eta_j$ and the transmission on the $j$-channel, i.e.
   \begin{equation}\label{etaT}
\eta_{true, j}=\frac{\eta_{j} }{\tau_{j}},
\end{equation}
with $j=s,i$. $\tau_{j}$ should be evaluated by means of an independent "classical" transmittance measurement. As in this paper
we present just a proof of principle of the proposed technique, we will not discuss this transmittance measurements anymore in
this paper. Thus, instead of providing the quantum efficiency of the CCD standalone, the results presented in the following can
be interpreted as the quantum efficiency of the whole optical system before the CCD, including the CCD itself.
\end{itemize}

\subsection{determination of CS}\label{evaluation of the CS}

\begin{figure}[tbp]
\par
\begin{center}
\includegraphics[angle=0, width=12cm, height=9 cm]{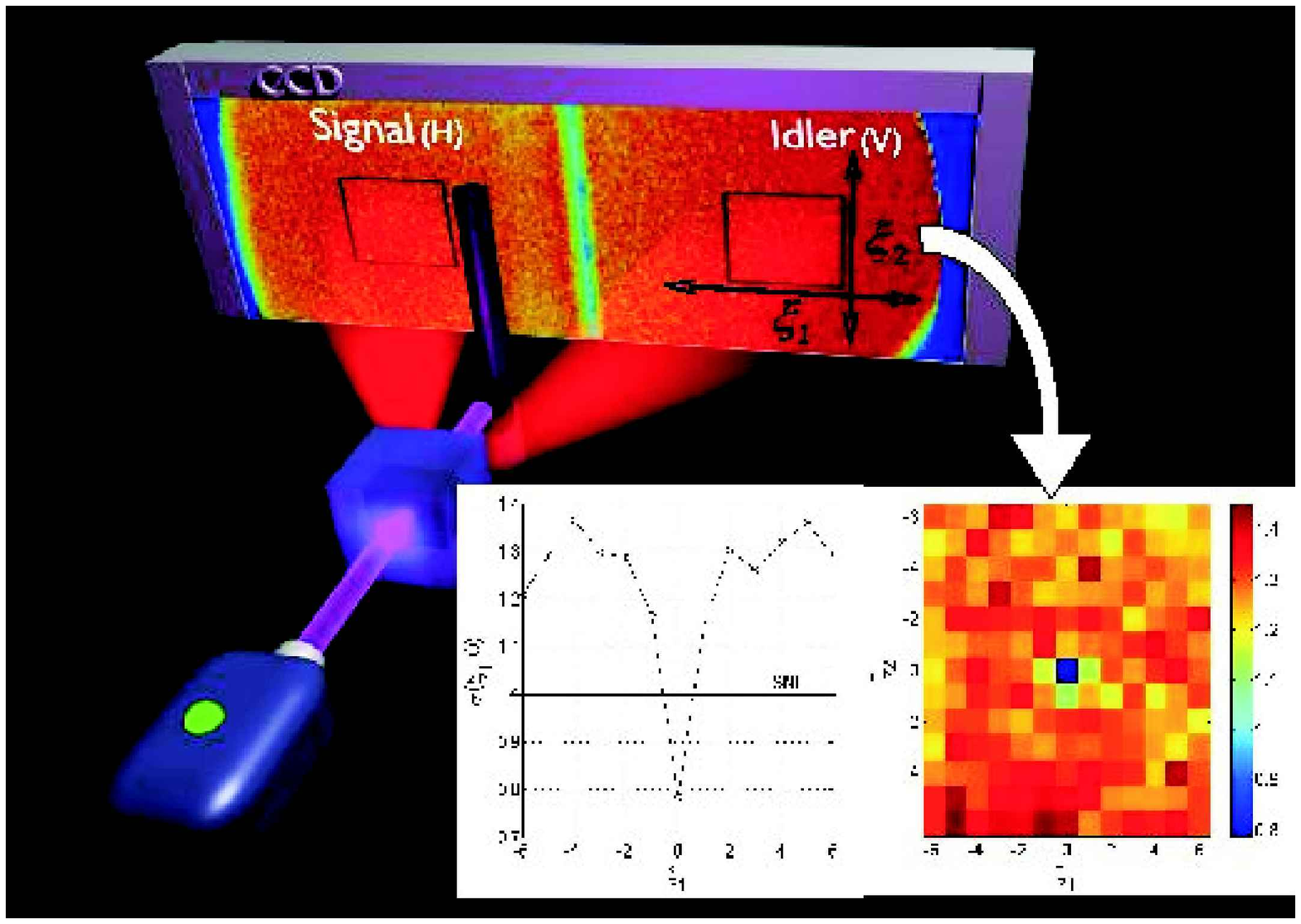}
\caption{\textsf{Determination of center of symmetry. The right hand side presents a typical image obtained by the CCD
in the working condition. The images is separated in two portions, one collecting the light from the signal beam (H-polarized)
and the other the light of the idler beam (V-polarized). Two regions $R_{s}$ and $R_{i}$ (squares) are selected, and subtracted
pixel by pixel. The spatial noise of the difference is reported in the inserts (right-bottom) in function of the position
$\mathbf{\xi}=(\xi_{1},\xi_{2})$ of the idler region.}} \label{CMdetermination}
\end{center}
\end{figure}
The single shot image is stored as a "superpixel" matrix (Fig. \ref{CMdetermination}). We select two equal rectangular regions
$\mathcal{A}_{det,s}$ and $\mathcal{A}_{det,i}$ belonging to the signal and idler portion of the image and containing a certain
number of "superpixels". In our case, we choose 9 $mm^2$ areas that span a wavelength bandwidth of the order of 10nm around the
degeneracy at $710nm$. Fixing the region $\mathcal{A}_{det,s}$, we evaluate the noise reduction factor $\sigma_{spatial}$ as a spatial average on the pairs of conjugated "superpixel" inside the two regions, in  function of the position of the center of the region $\mathcal{A}_{det,i}$ of
coordinate $\mathbf{\xi}=(\xi_1, \xi_2)$. In this way we obtain a matrix of values of the spatial average $\sigma_{spatial}(\xi)$ as a function of the
center $\mathbf{\xi}$ of the detection area $\mathcal{A}_{det,i}$.

The result obtained by the analysis of a typical image are presented in Fig. \ref{CMdetermination}, where a binning 6$\times$6
of the physical pixel has been applied and the number of photon per superpixel is $\simeq1700$. The presence of correlations
around $\mathbf{\xi}=0$ is represented by a deep in the values of the spatial average of $\sigma_{spatial}$, whereas far from the
minimum the correlation decrease, because conjugated pixels no more detect the correlated photons. Thus, this measurement
of the spatial correlation allows to determine the best position of $\mathcal{A}_{det,i}$ with respect to $\mathcal{A}_{det,s}$,
and hence the center of symmetry of the correlation. The size of the correlation deep represents the coherence area that in our experiment
is $\mathcal{A}_{coh}\sim 120\times120(\mu m)^{2}$.

The superpixel size used for the measurement must realize a tradeoff between the visibility of the correlation, and the final
uncertainty in the CS determination. On one side, as discussed in Sec.\ref{Multimode correlation}, for a good visibility of the
quantum correlation (and to increase the signal to electronic noise ratio) the "superpixel" area should be much greater than
$A_{coh}$. On the other side, a small pixel size will lead to a small uncertainty. We found that in our setup the best choice is
a pixel size equal to the coherence area, that can be obtained by a 6$\times$6 binning, as in Fig. \ref{CMdetermination}.

From the discussion above, it is clear that the measurement described allows a positioning of the two correlated (symmetrical)
regions within a single "superpixel" uncertainty, while the center of symmetry is identify within half a "superpixel". However,
it can be demonstrated that even a shift of a small fraction of "superpixel" with respect the real CS determine a increasing of
the noise reduction factor \cite{BrambPRA2004,BrambPRA2008}. In practice, the optimization of the NRF by micro positioning of
the CCD allows to determine the physical center of symmetry within a final uncertainty less than 1/10 of the "superpixel",
hence of the coherence area in our setup \cite{BrambPRA2004,BrambPRA2008}.

\subsection{Determination of the minimum size of $\mathcal{A}_{det,j}$}\label{A_det aetimation}

In the previous section we showed that CS can be determined with good precision, and the regions $\mathcal{A}_{det,s}$ and
$\mathcal{A}_{det,i}$ are consequently fixed to be strictly symmetrical and correlated even "locally" (i.e. for each pair of
twin spatial modes).

\begin{figure}[tbp]
\par
\begin{center}
\includegraphics[angle=0, width=12cm, height=7 cm]{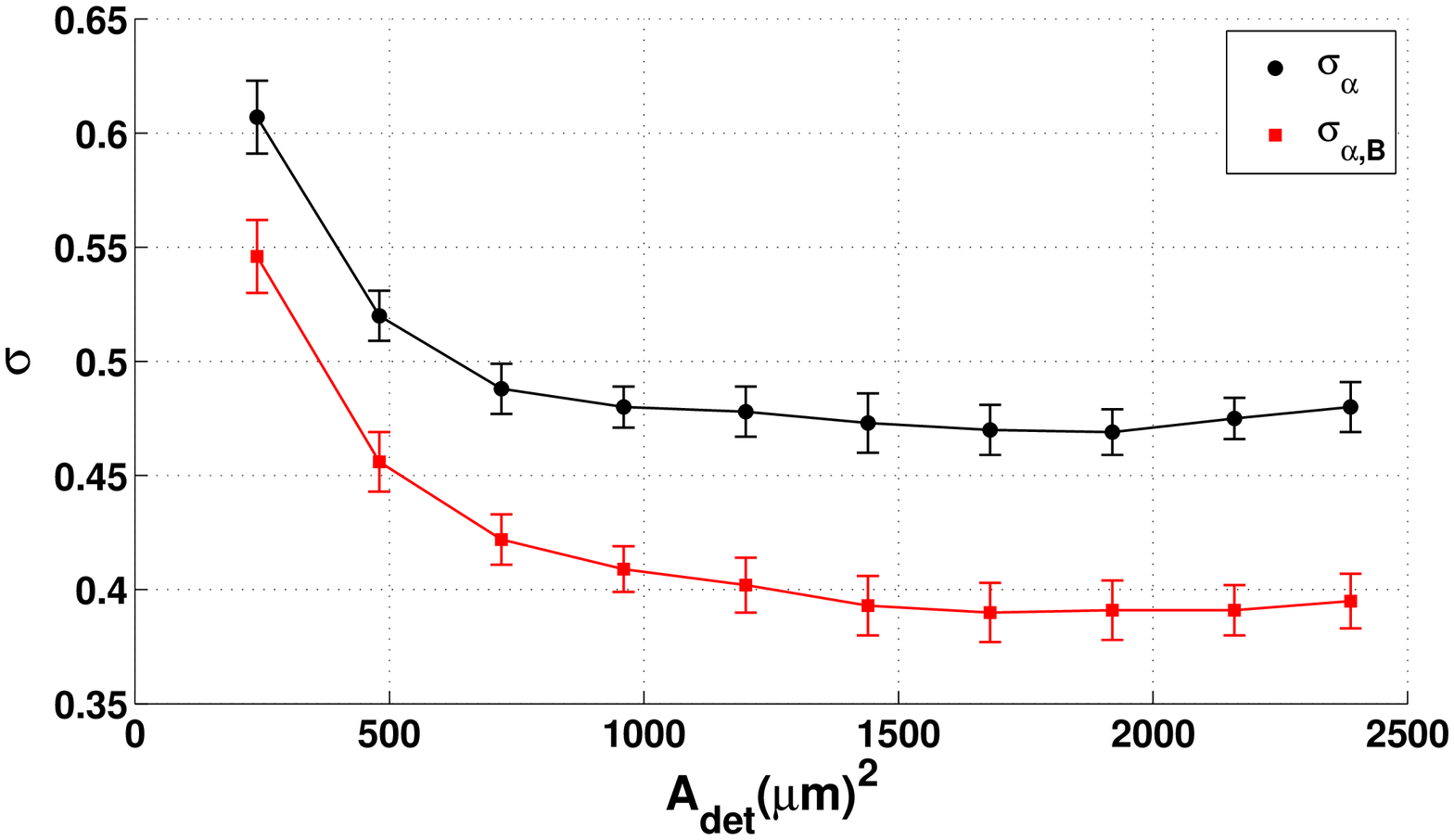}
\caption{\textsf{Noise reduction factor in function of the detection area. x-axis reports the linear size of the detection areas in $\mu$m. The two experimental curves refer to the noise reduction factor $\sigma_{\alpha}$ with (red), and  without (black) background correction.}} \label{SigmaVsAdetection}
\end{center}
\end{figure}

However, as pointed out in Sec. \ref{Multimode correlation}, perfect detection of quantum correlation requires the condition
$\mathcal{M}_{spatial}=\mathcal{A}_{det,j}/\mathcal{A}_{coh}\gg1$. In order to establish when this occurs, we measure the noise
reduction factor in function of the detection area $\sigma_{\alpha}(A_{det})$, i.e. in function of the number of spatial modes
detected. The measurement has been performed on a set of $\mathcal{N}=4000$ images using a binning $12\times12$ that means a
superpixel size of $240\times240(\mu m)^{2}$. Here, we define $N_{j}(k)$ the total number of photons detected in the region
$\mathcal{A}_{det,j}$ of the $k$-th image of the set. $\sigma_{\alpha}(\mathcal{A}_{det,j})$ has been evaluated according to Eq.
(\ref{sigmaalfaE}), over the set of 4000 images. Here, the single determination of the noise reduction factor is obtained by
subtracting the total numbers of photons detected in the two large regions $\mathcal{A}_{det,j}$ of the same image, and the
statistics is obtained over the set of images. The results are reported in Fig. \ref{SigmaVsAdetection}.

As expected, $\sigma_{\alpha}(\mathcal{A}_{det,j}$) is a decreasing function reaching an asymptotic value for
$\mathcal{A}_{det,j}>\mathcal{A}_{0}=1440\times1440(\mu m)^{2}$ that corresponds roughly to a number of spatial modes larger
than 150. Therefore we can affirm that, working with detection area larger than $\mathcal{A}_{0}$ allows to match the condition
$\mathcal{M}_{spatial}=\mathcal{A}_{det,j}/\mathcal{A}_{coh}\gg1$. This prevents possible bias in the estimation of quantum
efficiency due to the uncertainty in the propagation direction of correlated photons.

\subsection{Experimental contributions to the excess noise}\label{Contributions to the excess noise}

Other sources of experimental noise lead to systematic issues on the experimental values of
$\sigma_{\alpha}$ and consequently on the quantum efficiency obtained by Eq. (\ref{sigma_alfa}). In
the following we will address this problems.

The largest contribution to the excess noise that we have in our setup is related to the
instability of the Q-Switch laser pulse. In particular, we observed a fluctuation of the energy
pulse-to-pulse of more than $10\%$. The power $P$ of the pulse is directly related to the mean
value of photons per mode $\mu\propto\sinh^{2}(const*\sqrt{P})$. Therefore, the temporal statistics
of the PDC emission is drastically influenced by the pump power fluctuation. In particular $\mu$ is
not a constant. A temporal statistics on many pulses (many images) will be characterize by a mean
value $\overline{\mu}$ and variance $V\left(\mu \right)$. It can be demonstrated that the
contribution of the pump fluctuation modifies the expected value of the noise reduction factor with
respect to Eq.(\ref{sigma}) in the form
\begin{equation}\label{laser fluct}
\sigma\equiv\frac{\left\langle\delta ^{2}N_{-}\right\rangle} {\left\langle
N_{s}+N_{i}\right\rangle}=
1-\eta_{+}+\frac{\eta_{-}^{2}}{2\eta_{+}}\left[\overline{\mu}+\frac{1}{2}+
\frac{V\left(\mu\right)}{\overline{\mu}}\left(1+\mathcal{M}_{tot} \right)\right].
\end{equation}
From this equation it is clear that the instability of the pump generates a contribution to the
amount of excess noise that can be relevant, since it include a factor $\mathcal{M}_{tot}$.
\footnote{The order of magnitude in our experiment is estimable as large as
$2\eta_{+}\mathcal{M}_{tot} V\left(\mu\right)/\overline{\mu}=V\left(\langle
N_{s}+N_{i}\rangle\right)/\overline{\langle N_{s}+N_{i}\rangle}\sim5\cdot 10^{3}$ (see Tab.
\ref{Tab}, first two columns). It is four order of magnitude larger than the excess noise due to
the thermal fluctuations.} Following the same argumentation of Section \ref{Multimode correlation},
the effect of this enhanced excess noise can be suppressed if the losses on the two beams are a
posteriori compensated, by evaluating $\sigma_{\alpha}$ instead of $\sigma$. In this case
$\sigma_{\alpha}$ reduces again to Eq. (\ref{sigma_alfa}).

An other important source of excess noise is the background generated by the electronics  of the
CCD (digitalization, dark counts) and from the straylight, mostly caused by the fluorescence of the
filter and mirrors used for stopping the pump, and residual of the pump itself. The first
contribution, the electronic one, depend on the level of binning, and can be considered independent
with respect the thermal noise of the PDC, and straylight noise. In principle, also the straylight
noise is uncorrelated with respect the thermal fluctuation of the PDC light, and can be represented
by a poissonian-like statistics. However, some correlation between the straylight and the PDC
emission is introduced by the fluctuation of the pump, since when the pump pulse is more(less)
energetic both the PDC and straylight increase (decrease) accordingly. Anyway it can be
demonstrated that even this correlation cancel out in the difference of the photon number when the
transmission of signal and idler path are balanced. We define the number of counts $N_{s/i}'$
registered in the region $R_{s/i}$ expressed as the sum of the PDC photons $N_{s/i}$ and the
background $M_{s/i}$.

The expression linking the quantum efficiency to the expectation value of measurable quantities in
presence of background, the analogous of Eq. (\ref{sigma_alfa}), is

\begin{equation}\label{sigma_alfaB}
\sigma_{\alpha,B}\equiv\frac{\left\langle\delta
^{2}(N'_{s}-\alpha_{B}N'_{i})\right\rangle-\left\langle\delta
^{2}(M_{s}-\alpha_{B}M_{i})\right\rangle} {2\left(\left\langle N'_{s}\right\rangle-\left\langle
M_{s}\right\rangle\right)}=\frac{1}{2}(1+\alpha_{B})-\eta_{s},
\end{equation}
where $\alpha_{B}\equiv\frac{\eta_{s}}{\eta_{i}}=\frac{\langle N'_{s}\rangle-\langle
M_{s}\rangle}{\langle N'_{i}\rangle-\langle M_{i}\rangle}$. The background and its statistics can
be measured easily and independently, by collecting a set of $\mathcal{M}$ images when the PDC is
turned off, just by a $90^{o}$ rotation of the crystal. Following the same formalism of point d) of
the experimental procedure, $\alpha_{B}$ is obtained as:
\begin{equation}\label{alfaBE}
\alpha_{B}=\frac{E[N'_s (k)]-E[M_s (p)]}{E[N'_i (k)]-E[M_i (p)]},
\end{equation}
where $E[N'_j(k)]=\mathcal{N}^{-1} \sum_{k=1}^{\mathcal{N}}N'_j(k)$, and
$E[M_j(p)]=\mathcal{M}^{-1} \sum_{p=1}^{\mathcal{M}}M_j(p)$ represent the experimental
determination of the quantum expectation values $\langle N'_{j}\rangle$ and $\langle M_{j}\rangle$
respectively. At the same time,  $\sigma_{\alpha,B}$ in Eq. (\ref{sigma_alfaB}) is obtained by the
following experimental estimates:
\begin{eqnarray}\label{sigmaalfaBE}
\left\langle\delta^{2}(N'_{s}-\alpha_{B}N'_{i})\right\rangle\mapsto E[(N'_{s} (k)- \alpha_{B}
N'_{i} (k) )^2]-E[N'_{s} (k)- \alpha_{B} N'_{i} (k)]^2\\\nonumber
\left\langle\delta^{2}(M_{s}-\alpha_{B}M_{i})\right\rangle\mapsto
 E[(M_{s} (p)- \alpha_{B} M_{i} (p) )^2]-E[M_{s} (p)- \alpha M_{i} (p)]^2
\end{eqnarray}

We also mention that, for each measurement performed in the present work, the images affected by
cosmic rays have been discarded by a proper algorithm.

\subsection{Efficiency estimation and uncertainty evaluation }\label{Efficiency estimation and uncertainty}

In the previous Subsections we have presented the procedure to obtain an unbiased estimation of the
detection efficiency $\eta_s$ by means of appropriate positioning and sizing the detection regions,
a posteriori balancing of losses and background contribution analysis.  In this subsection we
present the experimental estimation of $\eta_s$ as well as its uncertainty budget.

The sizing parameter chosen for this experimental proof of principle
are: areas of detection $\mathcal{A}_{det,j}=2400\times3840(\mu
m)^{2}$ ($5\times8$ superpixels of size $480\times480(\mu m)^{2}$
obtained by a binning 24x24 of the physical pixels) corresponding to
about $640\times\mathcal{A}_{coh}$. The estimated value of $\eta_s$
is obtained by the inversion of Eq. (\ref{sigma_alfaB}), where
$\alpha_B$ is calculated according to Eq. (\ref{alfaBE}) and
$\sigma_{\alpha,B}$ performing the substitutions in Eq.s
(\ref{sigmaalfaBE})  on the basis of $\mathcal{N}$ images with PDC
light, and  $\mathcal{M}$ of background light. $\eta_i$ can be
obviously evaluated as $\eta_i=\alpha_B \eta_s$.

Once we intend to provide an uncertainty associated to the estimated value of $\eta_s$ (and
$\eta_i$), we should repeat the experiment $\mathcal{Z}$-times, i.e. we should collect
$\mathcal{Z}\cdot \mathcal{N}$ images with PDC light, and $\mathcal{Z}\cdot \mathcal{M}$ with
background light. Thus the estimated value of $\alpha_B$ and $\sigma_{\alpha,B}$ can be obtained
from
\begin{equation}\label{alfa^estim}
 \alpha|_{estim}=\mathcal{Z}^{-1}\sum^{\mathcal{Z}}_{l=1} \alpha(l) \qquad
 \sigma_{\alpha}|_{estim}={\mathcal{Z}}^{-1}\sum^{\mathcal{Z}}_{l=1} \sigma_{\alpha} (l)
\end{equation}
and the associated uncertainty is obtained following the guidelines of Ref. \cite{GUM}.
Specifically the uncertainty propagation is performed on the $2 \mathcal{N} + 2 \mathcal{M}$
measured quantities, namely $N'_s(k)$, $N'_i(k')$, $M_s(p)$, $M_i(p')$ with
$k,k'=1,...,\mathcal{N}$ and $p,p'=1,...,\mathcal{M}$. We accounted for the correlations between
$N'_s(k)$ and $N'_i(k')$ when $k=k'$, due mainly to PDC light, and between $M_s(p)$ and $M_i(p')$
when $p=p'$, due to pulse-to-pulse laser energy instability. Thus, it is assumed that there is no
correlation between measured quantities in different images. In our experiment
$\mathcal{N}=\mathcal{M}$ and $\mathcal{N}=500$ and $\mathcal{Z}=8$.

The estimated values of $\alpha_B$ and $\sigma_{\alpha,B}$ together with their uncertainties are
presented in Table 1, where, for the sake of comparison, are shown also the corresponding values
without background subtraction ($\alpha$ and $\sigma_{\alpha}$ respectively). According to Eq.
(\ref{sigma_alfaB}) we obtained $\eta_{s}=0.613 \pm 0.011$.

We underline that the estimated value of $\eta_s$ corresponds to the efficiency of the whole
quantum channel including the CCD -not only the efficiency of the CCD itself-, and that the
uncertainty associated to $\eta_s$ accounts only for the statistical contributions due to the
fluctuations of the $N$-s and $M$-s measured quantities (Type A uncertainty contributions,
according to Ref. \cite{GUM}).

Typically, further non-statistical uncertainties contributions (Type B \cite{GUM}) should be
accounted for in a complete uncertainty budget. For example, in this case a proper evaluation of
$\alpha_B$ with small uncertainty is mandatory to nullify the effect of the excess noise in Eq.
(\ref{laser fluct}), i.e. to ensure the validity of Eq. (\ref{sigma_alfaB}). A Type B uncertainty
contribution associated to the nullification of excess noise term should be considered. In Section
$\ref{Contributions to the excess noise}$ we showed that the excess noise due to the pulse to pulse
instability is of the order of $5\cdot10^{3}$ (see Footnote). From Eq. (\ref{laser fluct}) it turns
out that the condition for neglecting that term is
$\eta_{-}^{2}/(4\eta_{+}^{2})\cdot5\cdot10^{3}\ll1$, that in our case means $\eta_{-}\ll10^{-2}$.
The  uncertainty on $\alpha_{B}$ presented in table \ref{Tab}, shows that the balancing of the two
arms can be performed within $10^{-5}$, equivalent to the condition $\eta_{-}=10^{-5}$. Thus, the
possible contribution to the uncertainty due to this term is less than $10^{-6}$.

\begin{table}
  \centering
  \small
\begin{tabular}{||c|c|c|c|c|c|c|c|c||} \hline\hline
  $E[N'_s]$ & $\sqrt{E[\delta^{2} N'_s]}$ & $E[M_s]$ & $\sqrt{E[\delta^{2} M_s]}$ & $\alpha^{esim}$ & $\alpha^{estim}_{B}$ & $\sigma^{estim}$  & $\sigma^{estim}_{\alpha}$ & $\sigma^{estim}_{\alpha,B}$\\\hline\hline
   262710 & 35982 & 12751 & 1318 & 0.99952 & 0.99416 & 0.454 & 0.449 & 0.384 \\
    (620)   &  (437)  & (158) & (30) & (0.00003) & (0.00004) &(0.010) &(0.010) & (0.011)\\\hline\hline
\end{tabular}
  \caption{\textsf{Experimental estimates (row 2) and their uncertainties (row 3). Column 1-2 show the mean counts in the detection area $\mathcal{A}_{det,s}$ of the single images and the mean square of fluctuation $\delta^{2} N'_s(k)= (N'_s(k)-E[N'_s])^{2}$. Column 3-4 report the same values related to the background images. Column 5-6 present the experimental values of the compensation factor $\alpha$ and the value corrected for background counts. In column 7-9 are reported the raw noise reduction factor, the one after compensation of the losses, and finally the one after compensation and background correction respectively.}}\label{Tab}
\end{table}

In the determination of $\sigma_{\alpha,B}$ this uncertainty contribution is several order of
magnitude below the statistical (Type A) contributions, thus, absolutely negligible in the complete
uncertainty budget. A larger Type B contribution comes from the bias in the determination of the center of symmetry that in our present experiment is 1/10 of the coherence area in both the coordinates in the detection plane. Following the argumentation at the end of section \ref{evaluation of the CS} it generates an uncertainty of 1.5\%.
The actual relative uncertainty in the estimation of $\eta_s$ is $2.3\%$, it is
expected that it can be easily reduced of more than one order of magnitude increasing the number of
collected images, as far as the bottleneck of Type B uncertainty contribution is reached.

\section{Discussions and Conclusions}

As a test of consistency for the theoretical model at the basis of the proposed CCD calibration
technique, and, in particular, for the associated uncertainty model we evaluate the statistical
uncertainty associated to the mean values of $\alpha_B$ and $\sigma_{\alpha,B}$, according to
\begin{eqnarray}\label{sigma^estim}
 \Delta\alpha^{estim}=\sqrt{[\mathcal{Z}(\mathcal{Z}-1)]^{-1}\sum^{\mathcal{Z}}_{l=1}
 \left[\alpha(l)-\alpha^{estim}\right]^{2}}\\ \nonumber
 \Delta\sigma^{estim}_{\alpha}=\sqrt{[\mathcal{Z}(\mathcal{Z}-1)]^{-1}\sum^{\mathcal{Z}}_{l=1}
 \left[\sigma_{\alpha}(l)-\sigma^{estim}_{\alpha}\right] },
\end{eqnarray}
obtaining a good agreement with the estimated uncertainty.

\begin{figure}[tbp]
\par
\begin{center}
\includegraphics[angle=0, width=12cm, height=7 cm]{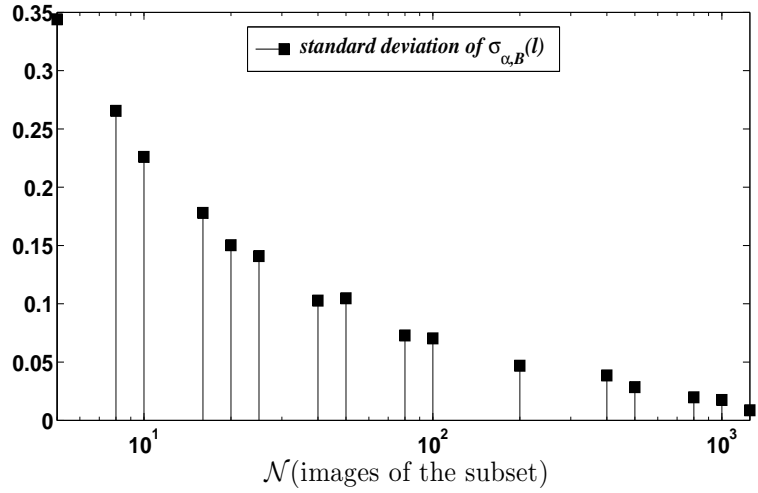}
\caption{\textsf{Standard deviation of $\sigma_{\alpha,B}$ on a number $\mathcal{Z}$} of
independent measurements. In each measurement a value of $\sigma_{\alpha,B}$ is obtained by formula
(\ref{sigma_alfaB}) with the substitutions (\ref{sigmaalfaBE}), using a set of $\mathcal{N}$
images. Although the standard deviation of the population decrease for larger $\mathcal{N}$, the
uncertainty on the mean value, only depends on the total number of images
$\mathcal{Z}\cdot\mathcal{N}$ [See text for further discussion].}\label{Std Sigma}
\end{center}
\end{figure}

Furthermore, we observe that according to the principle that the accuracy of a measurement depends
on the measuring time (in our case the number of acquired images), and not depend on how the data
are arranged, we verified that the final uncertainty on the mean values is not influenced by the
different possible choices of $\mathcal{Z}$ and $\mathcal{N}$, provided that total number of images
$\mathcal{Z}\cdot\mathcal{N}=const$. On the contrary, the standard deviation of the populations, is
a function of $\mathcal{N}$, as it is shown in Fig. \ref{Std Sigma}.

We note that Tab.\ref{Tab} shows that $\langle M_{s}\rangle$ is 5\% of the total counts $\langle
N'_{s}\rangle$ although the weigh of background correction is the 15\% of the estimated noise
reduction factor. Nevertheless, the uncertainty on the value of $\sigma_{\alpha,B}$ is just
slightly influenced by the background correction. For the sake of completeness we observe that the
electronic noise contributes with a standard deviation $\Delta\sim70$ counts. In a forward-looking
perspective, for pushing the uncertainty on this measurements at the level of the best values
obtained in the single photon counting regime, it would be very important to reduce the straylight.
Actually it is possible to design a different experimental configuration that limits the
fluorescence of the pump.

In conclusion we have proposed and demonstrated experimentally a new
method for the absolute calibration of analog detector based on
measuring the sub-shot-noise intensity correlation between two
branches of parametric down conversion. The results on the
calibration of a scientific CCD camera demonstrate the metrological
interest of the method, that could find various applications,
starting from the possibility to give a key element for redefining
candela unit \cite{qc} in terms of photo-counting.

\section*{Acknowledgments}
This work has been supported in part by MIUR (PRIN 2007FYETBY), NATO
(CBP.NR.NRCL.983251) and QuCandela EU project.

\end{document}